\documentclass[a4paper]{JHEP3}
\title{On the r\^ole of NLL corrections and Energy Conservation in the High Energy Evolution of QCD}
\author{Jeppe R.~Andersen\\Cavendish Laboratory, University of
  Cambridge\\JJ Thomson Avenue\\CB3 0HE\\Cambridge, UK\\Email:~\email{andersen@hep.phy.cam.ac.uk}}
\abstract{We present a new method for solving the BFKL evolution applicable
  at both leading and next-to-leading logarithmic accuracy, and tailored to
  the study of QCD multi-jet events at colliders. We utilise this
  to discuss corrections to the standard analysis. There are known,
  large corrections from energy and momentum conservation. We show that,
  despite claims to the contrary in the literature, these are unrelated to
  the next-to-leading logarithmic corrections to the evolution kernel.}
\preprint{Cavendish-HEP/06/06}
\newcommand{\dpn}{\ensuremath{\mathrm{d}\mathcal{P}_n}}
\newcommand{\Fn}{\ensuremath{\mathcal{F}_n}}
\newcommand{\dki}{\ensuremath{\mathrm{d}\mathbf{k}_i}}
\newcommand{\kj}[1]{\ensuremath{\mathbf{k}_{#1}}}
\newcommand{\qj}[1]{\ensuremath{\mathbf{q}_{#1}}}
\newcommand{\dy}[1]{\ensuremath{\mathrm{d}y_{#1}}}
\usepackage{amsmath}
\usepackage{cite}
\usepackage{epsfig}
\usepackage{feynmp}
\begin{document}

\section{Introduction}
\label{sec:introduction}
One of the many immediate challenges for QCD is to provide a reliable
description of the multiple hard jet environment of the LHC. Besides posing a
very interesting problem in itself, the QCD dynamics will provide signals
similar to that of many sources of physics beyond the standard model, and so
is very important to understand in detail. An intriguing alternative to the
standard approach of calculating the production rate of a few hard partons by
fixed order perturbation theory is to use the framework arising from the BFKL
(Balitskii--Fadin--Kuraev--Lipatov) equation to calculate the emission of
gluons (and quarks at next-leading logarithmic accuracy) from the evolution
of an effective, Reggeized gluon (Reggeon) propagator. The starting point
here is the observation that for e.g.~$2\!\to\!2, 2\!\to\!3,\ldots$ gluon
scattering, the partonic cross section in the limit where the rapidity span
of the two leading gluons is increased is dominated by the contribution from
Feynman diagrams with a $t$-channel gluon exchange. This $t$-channel gluon
is then evolved according to the BFKL equation, and will emit partons
accordingly.  Starting from the $2\!\to\!2$--gluon exchange, the
$2\!\to\!2\!+\!n$ gluon scattering process can be calculated in the
limit of large rapidity spans $\Delta$, thanks to the Regge factorisation of
the colour octet exchange. Obviously, this means that the formalism is
relevant only if there is sufficient energy at colliders to have multiple
emissions spanning large rapidity intervals. In this high energy limit, the
partonic cross section ($p_a', p_b'\to p_a, \{p_i\}, p_b$) factorises as
\begin{align}
  \label{eq:partonicxsec}
  \mathrm{d}\hat\sigma(p_a,p_b) = \Gamma_a(\mathbf{p}_{a})\
  f(\mathbf{p}_{a},-\mathbf{p}_{b},\Delta)\ \Gamma_b(\mathbf{p}_{b}),
\end{align}
where $p_a,p_b$ is the momentum of the partons furthest apart in rapidity,
$\Gamma_{a,b}$ are the process dependent impact factors, describing here the
gluon--gluon--Reggeised-gluon--coupling, and
$f(\mathbf{p}_{a},-\mathbf{p}_{b},\Delta)$ is the process independent gluon
Green's function, which is the object evolving according to the BFKL
equation. We use bold face vectors to denote the transverse components. This
is the same gluon Green's function that would enter calculations of the
small-$x$ evolution of parton density functions. The evolution of the gluon
Green's function can of course be discussed generally without reference to
specific impact factors, and so the conclusions of this study should impact
all applications of the BFKL formalism for colour octet or inelastic studies.
In Eq.~\eqref{eq:partonicxsec}, the partonic cross section has been
integrated over any number of gluons emitted in this evolution, so only the
dependence on the momenta of the leading jets have been retained.  The
BFKL approach should give a good approximation to the full
$2\!\to\!2\!+\!n$ calculation, when the hard jets are well separated in
rapidity.

The BFKL equation governing the evolution of the $t$--channel Reggeised gluon
was solved iteratively in
Ref.\cite{Kwiecinski:1996fm,Schmidt:1997fg,Orr:1997im} using the evolution at
leading logarithmic accuracy\cite{Fadin:1975cb,Kuraev:1976ge,Kuraev:1977fs},
and more recently in Ref.\cite{Andersen:2003an,Andersen:2003wy} at full
next-to-leading logarithmic accuracy\cite{Fadin:1998py,Ciafaloni:1998gs}.
The present paper has three purposes. Firstly, to present a new method for
obtaining the QCD evolution according to the BFKL equation. Secondly, to
discuss the implications of this new formalism on our understanding of the
sources of corrections. And thirdly, to announce the availability of a
computer code that calculates the exclusive multi-parton production rate
expected at the LHC or Tevatron according to the BFKL evolution.

\section{Iterative Solution of the BFKL Equation}
\label{sec:solut-bfkl-equat}
Our starting point for the discussion is the fully inclusive BFKL equation
describing the evolution of the gluon Green's function $f({\bf k}_a,{\bf
  k}_b,\Delta)$
\begin{align}
\label{eq:BFKLeqn}
\omega \ f_\omega\! \left({\bf k}_a,{\bf k}_b\right) = \delta^{(2+2\epsilon)} 
\left({\bf k}_a-{\bf k}_b\right) + \int \mathrm{d}^{2+2\epsilon}{\bf k} ~
\mathcal{K}_\epsilon\!\left({\bf k}_a,{\bf k}+{\bf k}_a\right) \ f_\omega\!\left({\bf k}+{\bf k}_a,{\bf k}_b \right),
\end{align}
where $w$ is the Mellin-conjugated variable to $\Delta$, and the BFKL kernel
$\mathcal{K}_\epsilon\!\left(\mathbf{k}_i,\mathbf{k}_j\right)$ is presently
known to next-to-leading logarithmic accuracy, where the logarithm is $\ln(
s_{ij}/{|\mathbf{k}_i||\mathbf{k}_j|})$, and $s_{ij}$ the invariant mass of
partons $i$ and $j$. The solution to the evolution from a momentum
$\mathbf{k}_b$ at rapidity $y_b$ to $\mathbf{k}_a$ at rapidity $y_a$
according to this integral equation can be written on the form
\begin{align}
\label{solution1}
f({\bf k}_a ,{\bf k}_b, \Delta) 
&=\sum_{n=0}^{\infty}\int\dpn\ \Fn,\nonumber\\
\int\dpn &= \left (\int\prod_{i=1}^n\dki\ \int_0^{y_0}\!\!\!\dy 1\int_0^{y_1}\!\!\!\!\!\!\dy 2\ \cdots\ \int_0^{y_{n-1}}\!\!\!\!\!\!\dy n\ \right)\ \delta^{(2)} \left({\bf
    k}_a + \sum_{l=1}^{n}{\bf k}_l - {\bf k}_b \right)\\
\Fn&=\left(\prod_{i=1}^n e^{\omega(\qj{i})(y_{i-1}-y_i)}\ V(\qj{i},\qj{i+1})\right)\ e^{\omega(\qj{n+1})(y_n-y_{n+1})}\nonumber
\end{align}
with $y_0=y_a\equiv\Delta$, $y_b=y_{n+1}=0$, $\qj{i}=\kj{a} +
\sum_{l=1}^{i-1}\kj{l}$, and, crucially, the real production vertices
$V(\qj{i},\qj{i+1})$ and trajectories $\omega(\qj{i})(y_{i-1}-y_i)$ are
regularised and finite at
LL\cite{Kwiecinski:1996fm,Schmidt:1997fg,Orr:1997im} and
NLL\cite{Andersen:2003an,Andersen:2003wy} in order to facilitate a direct
numerical evaluation. The correctness of the procedure at NLL was proved in
Ref.\cite{Andersen:2004uj} by comparing to analytic results for the evolution
in $\mathcal{N}\!=\!4$ Super Yang-Mills theory, since the analytic solution
to the BFKL equation is only known for conformal invariant theories (please
see Ref.\cite{Andersen:2005jr} for a discussion of the analytic methods at
NLL).

The formulation of the solution to the BFKL evolution in
Eq.~\eqref{solution1} allows in principle for the calculation of the fully
exclusive $2\!\to\!2\!+\!n$ cross section, i.e.~the expression of the
differential cross section in Eq.~\eqref{eq:partonicxsec} as
$\mathrm{d}\hat\sigma(p_a,\{p_i\},p_b)$. It is reassuring that this picture
is fully consistent with recent results on the multi-Regge form of QCD
amplitudes at the next-to-leading logarithmic
level\cite{Fadin:2003xs,Fadin:2003av,Fadin:2006bj}.
\begin{figure}
  \centering
  \begin{fmffile}{Phys_Int}
    \begin{fmfgraph*}(200,70)
      \fmfset{arrow_len}{3mm}
      \fmfset{arrow_ang}{15}
      \fmfpen{1pt}
      \fmfstraight 
      \fmfbottom{fi1,fi2}
      \fmftop{fo1,g1,g3,g4,g2,fo2} 
      \fmf{gluon,fore=black}{fi1,vup,fo1}
      \fmf{gluon,fore=black}{fi2,vul,fo2} 
      \fmffreeze
      \fmf{zigzag}{vul,vg2,vg3}
      \fmf{dots}{vg3,vg4}
      \fmf{zigzag}{vg4,vg1,vup}
      \fmffreeze \fmf{gluon,fore=black}{vg1,g1} \fmf{gluon,fore=black}{vg2,g2} 
      \fmflabel{{\small $k_a,\Delta=y_0$}}{fo2}
      \fmflabel{{\small $k_b, y_{n+1}=0$}}{fo1}
      \fmflabel{{\small $k_1, y_1$}}{g2}
      \fmflabel{{\small $k_n, y_n$}}{g1}
      \fmfv{f=black,d.sh=circle,d.si=0.07w,d.filled=shaded}{vul,vup}
      \fmfv{f=black,d.sh=circle,d.si=0.07w,d.filled=hatched}{vg2,vg1}
    \end{fmfgraph*}
  \end{fmffile}
  \caption{The $2\!\to\!2\!+\!n$ gluon scattering process described using
    Regge factorisation, with the initial state at the bottom. The shaded
    blobs are the gluon-gluon--Reggeon impact factors $\Gamma_{a,b}$, and the
    hatched blobs are the (regularised) gluon-Reggeon-Reggeon vertices. The
    Reggeized gluon (Reggeon) propagators are marked with zigzag lines. Gluon
    emission is generated in the rapidity span between the impact factors by
    the evolution described by the BFKL equation of the Reggeized gluon. At
    NLL the vertices can emit one or two gluons, or a quark-anti-quark pair.}
  \label{fig:ReggePicture}
\end{figure}
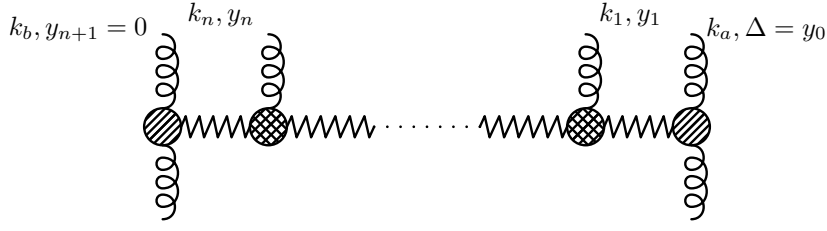

\section{Direct Solution of the BFKL Evolution}
\label{sec:direct-solution-bfkl}
Starting from Eq.~\eqref{solution1} it is possible to construct a new method
of evolving according to the BFKL kernel by first performing a simple change
of variables in the nested integration over rapidities, followed by the
introduction of another integral and delta functional, in order to make the
integration over rapidity separations independent
\begin{align}
  \label{eq:solution2}
  &\int_0^{y_0}\!\!\!\dy 1\int_0^{\dy 1}\!\!\!\!\!\!\dy 2\ \cdots\
  \int_0^{\dy
    {n-1}}\!\!\!\!\!\!\dy n\ \left(\prod_{i=1}^n e^{\omega(\qj{i})(y_{i-1}-y_i)}\right)e^{\omega(\qj{n+1})(y_n-y_{n+1})}\\
    =&\int_0^\Delta\mathrm{d}\delta y_n\int_0^{\Delta-y_n}\mathrm{d}\delta
    y_{n-1}\cdots \int_0^{\Delta-y_n-\cdots-y_2}\mathrm{d}\delta y_1\
    \left(\prod_{i=1}^{n} e^{\omega(\qj{i})\delta y_i}\right)
    e^{\omega(\qj{n+1})\delta y_{n+1}}\\
=&\int_0^\infty\mathrm{d}\delta y_{n+1}\int_0^\infty\mathrm{d}\delta
y_{n}\cdots\int_0^\infty\mathrm{d}\delta y_1
\ \delta(\Delta-\sum_{i=1}^{n+1}\delta y_i)\ \prod_{i=1}^{n+1} e^{\omega(\qj{i})\delta y_i},
\end{align}
so the gluon Green's function $f({\bf k}_a ,{\bf k}_b, \Delta)$ in
Eq.~\eqref{solution1} can be written as
\begin{align}
  \label{eq:solution3}
  f({\bf k}_a ,{\bf k}_b, \Delta) 
&=\sum_{n=0}^{\infty}\int\dpn\ \Fn,\nonumber\\
\int\dpn &= \left (\prod_{i=1}^n\int\dki\ \int_0^\infty \mathrm{d}\delta y_i
\right)\int_0^\infty\mathrm{d}\delta y_{n+1}\ \delta^{(2)} \left({\bf
    k}_a + \sum_{l=1}^{n}{\bf k}_l - {\bf k}_b \right)\delta\left(\Delta-\sum_{i=1}^{n+1}\delta y_i\right)\\
\Fn&=\left(\prod_{i=1}^n e^{\omega(\qj{i})\delta y_i}\
  V(\qj{i},\qj{i+1})\right)\ e^{\omega(\qj{n+1})\delta y_{n+1}}\nonumber
\end{align}
Using this, we see that $f({\bf k}_a ,{\bf k}_b, \Delta)$ is simply the value
at $\Delta\equiv\sum_{i=1}^{n+1} \delta y_i$ of the phase space integral of the
product of real--emission vertices $V(\qj{i},\qj{i+1})$ at rapidity
$y_i=\sum_{j=1}^i\delta y_j$ connected with Regge factors
$e^{\omega(\qj{i})\delta y_i}$ describing the probability of no (resolved)
emission between two adjacent (in rapidity) vertices\footnote{This interpretation
holds for $\omega(\qj{i})<0$, which can always be fulfilled in the
regularisation procedure of
Ref.\cite{Kwiecinski:1996fm,Schmidt:1997fg,Orr:1997im} at LL and
Ref.\cite{Andersen:2003an,Andersen:2003wy} at NLL.}. In
this case, it is clear that the \emph{virtual and unresolved} corrections
encoded in $\omega(\qj{i})$ lead to a \emph{decrease} with increasing
$\Delta$, while any \emph{increase} in $f({\bf k}_a ,{\bf k}_b, \Delta)$ is
due to the \emph{integration over phase space of the resolved emission} from
the vertices $V(\qj{i},\qj{i+1})$. 
The BFKL evolution can then be found by the following algorithm :-
\begin{enumerate}
\item Choose a random number of vertices for the evolution, $n\ge 0$
\item Generate a set $\{\mathbf{k}_i\}_{i=1,\ldots,n}$ of transverse momenta
  (the outgoing momenta are $\{-\mathbf{k}_i\}_{i=1,\ldots,n}$)
\item Calculate the corresponding set of trajectories
  $\{\omega(\qj{i})\}_{i=1,\ldots,n+1}$, and vertex factors\\
  $\{V(\qj{i},\qj{i+1})\}_{i=1,\ldots,n}$,
  $\qj{i}=k_a+\sum_{l=1}^{i-1}\kj{l}$ 
\item Generate the inter-vertex rapidity separations $\{\delta y_{i}\}$ according to the
  distributions $e^{\omega(\qj{i})\delta y_{i}}$
\item Calculate the corresponding $\Delta=\sum_{i=1}^{n+1}\delta y_i$ and
  return $\prod_{i=1}^n V(\qj{i},\qj{i+1})$
\item Repeat until required Monte Carlo accuracy is obtained
\end{enumerate}
This algorithm is vastly superior to the immediate Monte Carlo implementation
of the phase space integrals in Eq.~\eqref{solution1}. This is because that
instead of calculating, for a given $\Delta$, the contribution from any
number of emissions and any momentum configuration of these, this new method
calculates a representative rapidity span for a given number of emissions and
their configurations of momenta. Furthermore, this method is explicitly
symmetric in $k_a\leftrightarrow k_b$ and evolution direction (increasing or
decreasing rapidities), and it offers a more direct way of implementing the
necessary study at colliders of any length of the rapidity interval
$\Delta$, since this is automatically achieved by just a single sum over any
number of emissions. Unweighting of the Monte Carlo is also significantly
more efficient, since part of the integrand in the previous formulation
contributing to the variance of the integrand is now used to determine a
representative value for $\Delta$. We stress that both approaches provide
the same correct solution to the BFKL evolution.

\section{Phase space restrictions from energy and momentum conservation}
\label{sec:phase-space-restr}

Besides leading to a more efficient implementation of the BFKL evolution, the
new formulation of the solution to the BFKL evolution is also useful for
discussing sources of corrections to the standard BFKL analysis. The phase
space restrictions from energy and momentum conservation can be implemented
simply by adding a step in the algorithm rejecting events prohibited by phase
space considerations. We see that the discussion of the region of integration
for real emission is completely independent of the logarithmic accuracy of
the evolution kernel expressed in terms of the vertices $V(\qj{i},\qj{i+1})$
and Regge trajectories $\omega(\qj{i})$.  This clearly shows that, despite
what is often claimed in the literature (see
e.g.~Ref.\cite{Kwiecinski:2000bd} and references therein), phase space
restrictions in terms of e.g.~energy and momentum conservation are completely
independent of the logarithmic accuracy to which the BFKL evolution is
performed.  Specifically, the NLL corrections to the evolution kernel
\emph{do not} implement energy and momentum conservation.  Furthermore, the
standard solution of the LL and guestimate of the NLL evolution based on a
Mellin transform in the transverse momentum of the kernel will \emph{always
  fail} to take such considerations into account, by its very nature as fully
inclusive both in number of emissions and as an integral to infinity of the
transverse momentum. This is simply a result of the fact that the BFKL
evolution is local, while energy and momentum conservation depends on the
full final state configuration. Such considerations should be implemented by
modifications to the evolution beyond the discussion of the logarithmic
accuracy of the evolution. Please note that this discussion of energy and
momentum conservation goes beyond the discussion of longitudinal momentum in
collinear splittings at small-$x$, implemented by ensuring the vanishing of
the first moment (see e.g.~Ref.~\cite{Ciafaloni:1999yw}).

While it is not our job to guess the cause of the confusion in the literature
over the r\^ole of energy and momentum conservation and higher logarithmic
corrections to the evolution kernel, it is perhaps beneficial to discuss the
differences and similarities between the multi-particle generating
colour-octet exchange and the colour-singlet exchange relevant for
diffractive studies. The evolution in rapidity $\Delta$ of the gluon Green's
function for both cases is described by a BFKL equation. In the case of
diffractive $2\to 2$ processes, the rapidity span of the BFKL evolution is
given by $\Delta\approx\ln s/{s_0}$, when the Regge scale $s_0=|\mathbf{k}_a|
|\mathbf{k}_b|$ is the product of the transverse momentum of the two
(massless) scattered particles.  However, the impact factors depend on $s_0$
only at NLL accuracy, and so it can be argued that only at NLL accuracy does
the prediction gain a correct dependence on the centre of mass energy
(although of course the evolution can (and should!) be discussed without
reference to specific impact factors).  While this is true, care has to be
taken when discussing instead colour-octet exchange with multiple emissions.
First of all, there is no one-to-one correspondence between the rapidity span
of the evolution and the centre of mass energy, although obviously the centre
of mass energy tends to increase with increasing rapidity span.  Note
specifically that if one sets $s=s_0 e^\Delta$ with the Regge scale
$s_0=|\mathbf{k}_a| |\mathbf{k}_b|$, then any additional emission from the
BFKL evolution is kinematically excluded. The error\footnote{It is often
  argued that this correction is logarithmically subleading. However, as we
  have demonstrated this is not part of the subleading corrections to the
  evolution kernel.} in ignoring the contribution from the BFKL emission to
the centre of mass energy is significant\cite{Andersen:2003gs,Avsar:2005iz}.
Secondly, any constraint on the real emission will significantly lower the
expected BFKL signatures, e.g.~the expected rise in $F_2$ at small-$x$, and
the increasing jet azimuthal decorrelation with rapidity at hadron colliders.
In fact, the effect of correctly implementing energy and momentum
conservation on top of the LL BFKL evolution has a larger impact on the jet
azimuthal decorrelation even at LHC energies than the inclusion of the NLL
corrections to the evolution kernel (compare e.g.~Fig.~10 of
Ref.\cite{Andersen:2003wy} with Fig.~5 of Ref.\cite{Orr:1997im}).

\section{Conclusions}
\label{sec:conclusions}

We have presented a new method for solving the BFKL evolution very
effectively for the study of multi-partonic final states at hadron colliders.
We have furthermore demonstrated that the discussion of energy and momentum
conservation and phase space constraints of the evolution in general is
completely separate to the discussion of the logarithmic accuracy of the
evolution kernel, contrary to the claims found in the literature. We have
demonstrated how energy and momentum conservation can be implemented.

A program implementing the new algorithm for the BFKL evolution in the case
of multi-jet production at Tevatron and LHC energies is available at the URL\\
\verb!http://www.hep.phy.cam.ac.uk/~andersen/BFKL!\\
The current version implements energy and momentum conservation, and the
BFKL evolution kernel at LL supplemented by the running coupling terms from NLL.

\acknowledgments 
I would like to thank Einan Gardi, Albrecht Kyrieleis, Robert
S.~Thorne, Agust\'in Sabio Vera, and Bryan R.~Webber for useful
discussions.\\ 
This research was supported by PPARC (postdoctoral fellowship
PPA/P/S/2003/00281).

\bibliographystyle{JHEP}
\bibliography{database}

\end{document}